\documentclass{aa}  

\usepackage{graphicx}
\usepackage{txfonts}
\newcommand{\blaz}{1508+5714}
\newcommand{\dotM}{{\bf$\rm \dot M$}}

\begin{document}

   \title{NVSS J151002+570243: accretion and spin of a $z>4$ {\it Fermi} detected blazar}
    \titlerunning{1508+5714: accretion \& spin in early Universe}

   \author{G. Alzati
          \inst{1,2}
          \and
          T. Sbarrato
          \inst{3}
          \and
          G. Ghisellini
          \inst{3}
          }

   \institute{Dipartimento di Fisica ``Aldo Pontremoli", 
              Università degli Studi di Milano, 
              Via Celoria 16, 20133 Milano, Italy
         \and
             Dipartimento di Fisica, Sapienza Università di Roma, Piazzale Aldo Moro 5, 00185, Roma, Italy \\
             \email{alzati.2200689@studenti.uniroma1.it}
         \and
             INAF - Osservatorio Astronomico di Brera,
             via E. Bianchi 46, 23807 Merate (LC), Italy\\
             }

  \abstract{
Active galactic nuclei at high redshift feature two most peculiar characteristics connected to key open questions in AGN physics: they are overly massive, which is not explainable via sub-Eddington accretion, and most of their population hosts relativistic jets, that are commonly connected to spinning black holes and hence less effective in accreting mass onto the black hole.
The formation and evolution of such massive objects is currently an open puzzle.
NVSS J151002+570243 is part of this population, being 
the most distant blazar consistently detected by {\it Fermi}/LAT,
hence hosting a powerful jet. 
We tested the hypothesis of a super-Eddington accretion process for this source by modeling its big blue bump with a set of accretion disk emission models. 
We first tested a standard geometrically thin, optically thick $\alpha$-disk, obtaining a mass of Log$M/M_\odot=8.65\pm0.19$ consistent with virial-based results and a significantly sub-Eddington accretion rate $\lambda=0.02\pm0.01$. 
We then focused on the analytic approximations of two numerical models 
that take into account the General Relativity effects of a spinning black hole (reasonable due to the presence of a jet) and a close-to- or super-Eddington accretion rate (KERBB and SLIMBH).
Despite the focus on super-critical accretion, these models confirm a surprisingly low Eddington ratio, of the order of 3\%.
The hypothesis of a continuous accretion at this measured rate is unrealistic, since it would imply a seed black hole mass of $\sim10^6-10^8M_\odot$ at redshift $z=20$. 
Hence we explore the possibility of a continuous super-critical accretion starting from a $\sim10^2M_\odot$ seed, that would spin up the black hole and eventually contribute in launching the relativistic jet. 
The measured low accretion rate would thus happen only once the jet is active. 
This idea would reconcile the black holes with large masses accreting at somewhat slow rates that are observed at $z>4$, with the need of an extremely fast evolution, by allowing the formation of stellar-size black hole seeds even as late as at $z\sim8$. 
}

   \keywords{ Galaxies: active, Galaxies: high-redshift, Galaxies: jets, quasars: supermassive black holes, Methods: analytical
               }

   \maketitle

\section{Introduction}

The search of supermassive black holes (SMBHs) in the early Universe has led to the discovery of more than 400 quasars at $z>5.7$ in the last 10 years \citep[see for a review][and references therein]{fan23}, and it seems far from being complete. 
During this time, an always increasing fraction of jetted sources have been discovered at $z\geq4$, leading to the conclusion that relativistic jets are significantly more numerous among the observed quasar population at high redshift, compared to their analogous lower redshift counterparts \citep[see][for a review]{sbarrato21}.
The SMBH masses of this ever growing sample of sources are at the highest end of the AGN mass distribution: on average they show $M\sim10^9M_\odot$ or more. 
The evolution of such massive black holes requires time, and the age of the Universe at which they are observed does not provide enough time for a continuous accretion following standard assumptions ($z\sim6$ corresponds to a Universe age of 900Myrs, while $z\sim4$ is about 1.5Gyrs). 
This has been dubbed as the ``mass problem", and currently represent one of the most intriguing open questions of AGN physics.
Jets worsen the scenario, since they are typically associated to spinning black holes, that are thought to have larger radiative dissipation and thus slower effective mass accretion compared to non-spinning systems.

The excess of jetted sources has been studied by tracing their population via blazar discovery and characterization. 
As a matter of fact, following simple geometrical assumption, the number of jetted sources with specific features in a well defined redshift range can be estimated by counting their aligned counterparts. 
This approach is particularly efficient if a blazar is strictly defined as a source observed under a viewing angle $\theta_{\rm v}$ smaller than its beaming angle $\theta_{\rm b}$. 
The jet emitting region is strongly relativistic and we can assume that $\theta_{\rm b}\simeq\sin\theta_{\rm b}$, and thus $\theta_{\rm b}\sim1/\Gamma$. 
Assuming that self-similar AGN jets in a given volume are homogeneously distributed in their orientation, for each detected blazar with $\theta_{\rm v}<1/\Gamma$ one can assume the presence of $\sim2\Gamma^2$ analogous sources with their jets oriented randomly in the sky. 
We know from the literature that the bulk Lorentz factors of jet emitting regions in blazars are typically in the range $\Gamma\sim8-15$, when derived from a broad-band spectral energy distribution (SED) modeling that includes the high-energy component\footnote{
The highest values are derived for $\gamma$-ray detected sources included in the {\it Swift}/BAT and {\it Fermi}/LAT all-sky catalogs, while at $z>4$ the Lorentz factors appear to be slightly smaller. This might be due to an intrinsic difference, or even to a lack of $\gamma$-ray detection, that does not allow for a proper modelling of the high-energy SED component.
}.
Such values imply that for each observed blazar, $\sim130-450$ AGN with the same intrinsic features (e.g.\ SMBH mass and accretion rate, jet power, nuclear features) must exist at the same redshift, making blazars the most efficient tracers of the overall jetted population at high redshift.
At low redshift the jetted AGN population is easily traced with minor loss in completeness thanks to their extreme radio brightness, that allow jetted sources to be easily included in all-sky radio surveys. 
The intensified Cosmic Microwave Background (CMB) energy density (scaling as $(1+z)^4$) is instead responsible for quenching the extended radio emission at roughly $z>3.5$ \citep{ghisellini13}, not allowing an equivalently complete mapping of the jetted AGN population in the radio frequencies. 
Blazars and their high-energy observation and classification are crucial to trace relativistic jets in the early Universe.

The systematic search of blazars at $z>4$ lead to an interesting suggestion: the jetted AGN population traced with this approach is significantly more numerous than the jetted fraction observed in the lower redshift Universe \citep[see e.g.][]{sbarrato21}, that is commonly thought to be around 10\%, and the one derived at high-$z$ from radio observations \citep[$9.4\pm5.7\%$,][]{liu21}. 
The current $z>4$ blazar (and thus overall jetted AGN) population is dominated by $M>10^9M_\odot$, because of limiting sensitivities of all-sky surveys.
Therefore it has been concluded that at least the most massive active SMBHs in the early Universe have a much larger jetted fraction, and likely prefer to form and evolve in jetted AGN. 
This was already suggested by \cite{Volonteri11}, that highlighted a slight excess of jetted sources at $z>3$ by comparing their comoving number density extracted from the blazar luminosity function of the 3-year Burst Alert Telescope (BAT) all sky survey \citep{ajello09} with the Sloan Digital Sky Survey (SDSS) quasar luminosity function \citep{Shen2011}. 
Further confirmation was built up in the following years, up to the currently highest redshift blazar discovered \citep[VLASS J041009.05-013919.88;][]{banados24}, that extends at $z=7$ the issue of finding a significant excess of jetted sources. 

The main issue with an extended search of high-$z$ blazars is that the commonly used smoking gun of blazars nature, i.e.\ a $\gamma$-ray detection, is not as efficient at $z>3$: the all-sky catalogs of $\gamma$-ray sources provided by the Large Area Telescope (LAT) onboard the {\it Fermi} satellite \citep{atwood09} do not reach deep enough sensitivities to detect the steeply falling $\gamma$-ray spectra of blazars at such high redshift. 
A careful selection of blazar candidates with secure redshift estimates and possibly radio emission is necessary, followed by X-ray observations to confirm the presence of the intense, relativistically beamed inverse Compton emission typical of jets aligned close to our line-of-sight. 
Motivated by the abundance of jetted sources at high-$z$, the {\it Fermi}/LAT collaboration focused on the search of the most powerful known blazars in their data, spanning about 10 years of observations.
Following a dedicated data analysis, \cite{ackermann17} managed to confirm the detection of 5 blazars at $z>3.1$ in $\sim92$ months of {\it Fermi} Pass 8 source class photons in the 60 MeV to 300 GeV energy range.
Among them, NVSSJ151002+570243 ($z=4.31$) was at the time the highest-redshift $\gamma$-ray source  detected by {\it Fermi}/LAT. 
The five {\it Fermi} sources obtained with this approach are among the most extreme blazars in terms of jet power, due to their extreme Compton Dominance, and also host extremely massive accreting SMBHs ($M\sim10^9-10^{10}M_\odot$). 
They clearly contribute to tracing a significant excess of jetted sources in the high-$z$ Universe. 

This first systematic search of $z>3.1$ $\gamma$-ray blazars showed the {\it Fermi}/LAT capability of detecting such faint and distant sources, opening the way to other detections in the first 2 billions years of the Universe. 
To obtain significant detections, pointed searches in binned light curves were performed, to increase the {\it Fermi}/LAT data test statistics and catch possible blazar flaring events at high-$z$.
Following this approach, \citet{liao18} and \citet{kreter20} enlarged the sample of {\it Fermi}-detected high-$z$ blazars, extending it to $z=4.72$ with the discovery of B3~1428+422.

In this work we will focus on the continuously {\it Fermi}-detected $z>4$ blazar, in order to characterize its jet (Section \ref{sec:jet}) and accretion emission (Section \ref{sec:disc}). We aim at finding signs of fast accretion, that could help in justifying its presence in the first 1.5Gyrs of the Universe, and ultimately discuss how its jet might be involved in its fast  evolution (Section \ref{sec:salpeter}).
In the following we adopt a flat cosmology with a Hubble parameter of $H_{\rm 0} = 70\,{\rm km\,s^{-1}\,Mpc^{-1}}$, and cosmological density parameters $\Omega_{\rm M} = 0.3$ and $\Omega_{\rm \Lambda} = 0.7$. 

\section{\blaz: a powerful jet 1.5Gyrs after the Big Bang}
\label{sec:jet}

   \begin{figure}
   \centering
      \includegraphics[width=1.1\hsize]{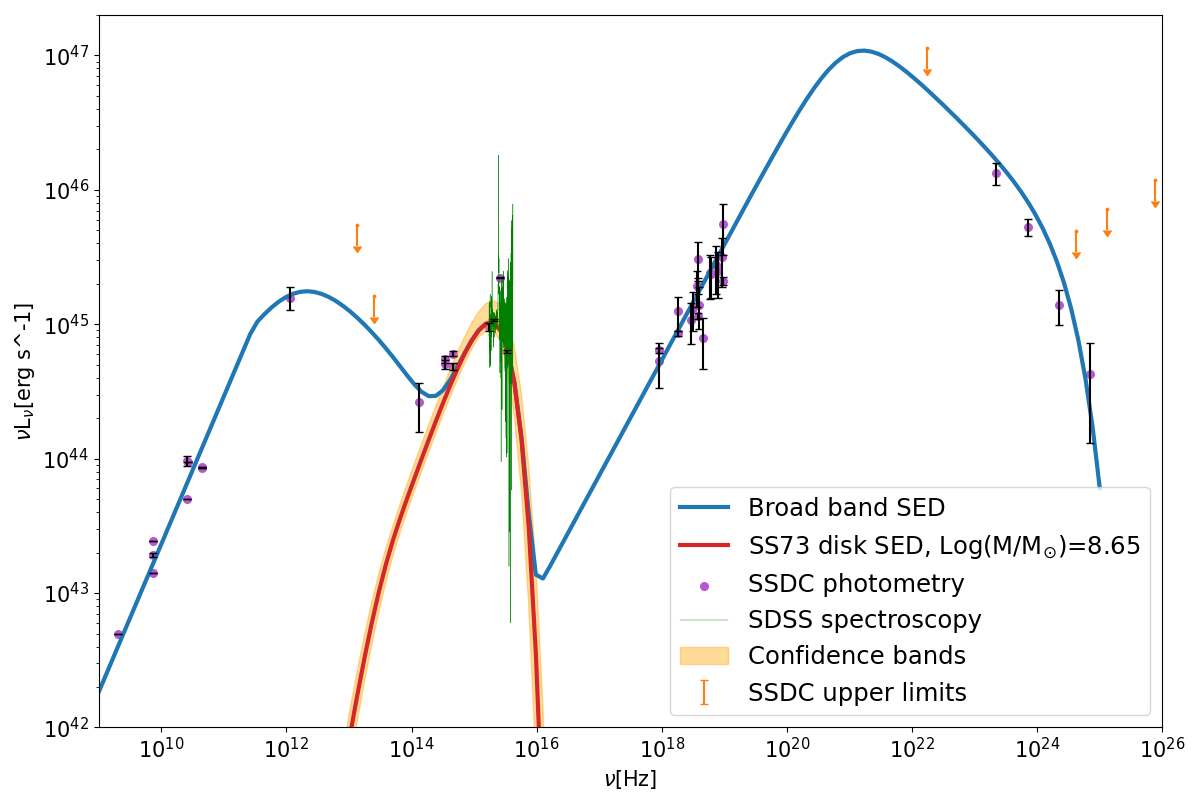}

      \caption{Broad band SED of \blaz. The blue line shows the total SED that best model the data. The SED includes the jet component (dominated by the two prominent humps in radio and at high-energies) and the accretion disk emission, shown with a red line and yellow band to show its confidence range. Photometric data are shown with purple dots and orange upper limits, while the green line shows the SDSS spectroscopic data.
      The data set and related best modelled SED clearly show the typical behaviour of a high-$z$ blazar, with flat radio spectrum, strong Compton dominance and prominent big blue bump.}
         \label{Fig:SED}
   \end{figure}

In order to test the hypothesis of a SMBH accreting at super-Eddington ratio, we studied the SED of NVSS J151002+570243 (also known as SDSS J151002.92+570243.3, that will be referred to as \blaz\ in the following), a high redshift blazar ($z=4.31$). 
Our goal was to estimate the physical parameters of its central black hole by comparing photometric data with three different accretion disk models, each based on different physical assumptions.
To this aim, we collected archival data from the Space Science Data Center (SSDC\footnote{\url{https://www.ssdc.asi.it}}) and the SDSS\footnote{\url{https://www.sdss.org}} \citep{Shen2011}.

We first focused on \blaz\ broad band SED, in order to characterize its relativistic jet emission.
To do so we adopted a phenomenological model presented in \cite{Ghisellini_Fermi_blazar_sequence} that approximates the SED produced by a relativistic jet aligned close to our line-of-sight with a set of smoothly broken power-laws \citep[for a realistic, physical motivated SED modeling refer to e.g.][]{gokus24}. 
The typical blazar SED is characterized by two humps, 
generally associated to synchrotron and inverse Compton emission, peaking respectively in the radio-IR and X-$\gamma$-rays. 
The overall emission profile can therefore be described by the following set of equations.
\begin{equation}
    L_R(\nu)=A\nu^{-\alpha_R};\:\:\:\:\:\nu\leq\nu_t
\end{equation}
that connects at $\nu_t$ with:
\begin{equation}
    L_{S+C}(\nu)=L_S(\nu)+L_C(\nu);\:\:\:\:\:\nu>\nu_t
    \label{eq_FBS_R}
\end{equation}
$L_S$ and $L_C$ are, respectively, the luminosities of the emission produced by the synchrotron radiation and inverse Compton scattering, which are computed as: 
\begin{equation}
    L_S(\nu)=B\frac{(\nu/\nu_S)^{-\alpha_1}}{1+(\nu/\nu_S)^{-\alpha_1+\alpha_2}}e^{-\nu/\nu_{cut, S}}
    \label{eq_FBS_S}
\end{equation}
\begin{equation}
    L_C(\nu)=C\frac{(\nu/\nu_C)^{-\alpha_3}}{1+(\nu/\nu_C)^{-\alpha_3+\alpha_2}}e^{-\nu/\nu_{cut, C}}
    \label{eq_FBS_C}
\end{equation}

A, B and C are constants, obtained by imposing and solving the following conditions:
\begin{itemize}
    \item $L_{S+C}(\nu_t)=L_R(\nu_t)$
\item $L_{S+C}(\nu_S)=L_S(\nu_S)$ with $\nu_S$ synchrotron peak frequency
\item $L_{S+C}(\nu_C)=L_C(\nu_C)$ with $\nu_C$ inverse Compton peak frequency
\end{itemize}

High-$z$ blazars are among the most powerful blazars known, but they show some differences in their SEDs with respect to the brightest sources in 2FGL (from which the phenomenological model is compiled). 

Generally, they show a less prominent synchrotron emission with respect to their optical luminosity \citep[always accretion dominated, see e.g.][]{sbarrato15}. 
A partial quenching of their radio emission by the large CMB energy density (that scales as $\propto(1+z)^4)$ is likely the main reason behind the radio faintness of $z>3$ blazars \citep{ghisellini15}.
This leads to a stronger Compton dominance, that does not necessarily translates in a different radiation field involved in the inverse Compton emission, when compared to lower-$z$ blazars.

Therefore, we adapted the parameters describing the highest luminosity bin in  \cite{Ghisellini_Fermi_blazar_sequence}, obtaining the values shown in Table \ref{Phenom_SED_Params}. 
The parameters describe an extremely powerful FSRQ, with more extreme features (higher CD, smaller synchrotron peak frequency) than the highest luminosity bin defined by Ghisellini and collaborators. This is consistent with what observed at $z>4$ \citep{sbarrato12,sbarrato13}, i.e.\ sources among the most powerful blazars known. Note that this is likely a selection effect, since we are still observing only the tip of the iceberg of the blazar population, especially in their high-energy emission.

Aside from classification purposes, characterizing the relativistic jet allowed us to evaluate its contribution to the optical/UV emission and therefore a more precise analysis on the accretion disk. 

\begin{table}
\caption{Parameters used in the phenomenological SED to describe the broad band photometry dataset}             
\label{Phenom_SED_Params}      
\centering          
\begin{tabular}{cc}
        \hline\hline
      $\alpha_1$   &  0.3 \\

        $\alpha_2$ & 1.45\\
     
        $\alpha_3$ & 0.14\\
       
        $\nu_t$ & $3\times10^{11}$Hz\\
        
        $\nu_S$ & $1.5\times10^{12}$Hz\\
      
        $\nu_C$ & $1\times 10^{21}$Hz\\

        $\nu_{cut, S}$& $2\times10^{14}$Hz\\
        
        $\nu_{cut, C}$& $2.5 \times 10^{24}$Hz\\
   
        $\nu_SL(\nu_S)$ &$1.76\times10^{45}\rm erg\: s^{-1}$\\

        CD & $60$\\
\hline
    \end{tabular}
\end{table}

\section{Accretion and black hole mass}
\label{sec:disc}

High-$z$ blazars clearly show ``big blue bumps" in their rest-frame optical-UV emission, similarly to bright quasars. This feature is generally associated with a radiatively efficient, geometrically thin, optically thick accretion disk. 
We therefore modeled this emission with three reasonable accretion disk emission models: the most simple, standard $\alpha$-disk model \citep{SS73}, and two models that take  into account General Relativity effects on the disk emission, along with the influence of the SMBH spin and possibly close-to- or super-Eddington accretion. 
We took advantage of the analytic approximations of the numerical models KERRBB \citep{KERRBB} and SLIMBH \citep{sadowski09}, that  \cite{Campitiello_18,Campitiello_19} developed in order to extend their applications to SMBHs and AGN in particular. 
The analytic approximations are particularly convenient, because they explicitly depend on the main physical parameters of the central SMBH, i.e.\ mass, accretion rate and spin, without being only linked to X-ray data analysis, as happens for stellar black holes.
We therefore used three different accretion disk emission models to derive independent estimates of \blaz\ mass $M$, spin $a$ and most importantly its accretion rate $\dot M$.

As an independent measure and comparison tool, we considered the virial black hole mass Log$\,M_{\rm vir}/M_\odot=8.52 \pm 0.39$ derived from the optical spectroscopic analysis by \cite{Shen2011}. This is the value obtained from the CIV broad emission line. 
While the three disk emission models are intertwined (see below for details), this is a completely independent mass measure method, and provides crucial information to study \blaz\ accretion features.

\subsection{A standard $\alpha$-disk}
\label{sec:alpha}
The first step of our accretion disk analysis focussed on characterizing its emission using the standard $\alpha$-disk model by \cite{SS73}, that in the following we will refer to as SS73. This model considers the accretion of matter on a non-rotating black hole in Newtonian approximation, therefore its emission profile depends only on the mass ($M$) of the central SMBH and its accretion rate ($\dot M$). 
We also considered the anisotropy due to the disk shape and the viewing angle, that is described by a factor $2\cos\theta$ \citep{calderone13}.
Blazars are observed face on, and thus we can assume $\theta\sim 0$:

\begin{equation}
    L_{\rm disk}(\theta)=2\cos\theta \dot Mc^2 =2\eta\dot Mc^2
\end{equation}
Note that the optical spectrum suffers from absorption due to intervening Ly$\alpha$ clouds at wavelengths smaller than $\lambda_{\rm Ly\alpha}=1216{\rm\AA}$. 
We first tried to correct the extended absorption due to the Gunn-Peterson effect via the \cite{meiksin06} algorithm, but this source appears 
significantly less absorbed than the overall $z>4$ population, and the corrected flux showed an uncommon and extreme slope, suggesting that we were over-de-absorbing it. 
For a safer approach, we thus decided to exclude the absorbed section of the spectrum from our analysis, even if this meant a loss of rest-frame UV information.

Photometric and spectroscopic data were then compared with the SS73 emission profile, to obtain estimates of mass and accretion rate:
\begin{equation}
\label{SS_results}
\begin{split}
    &{\rm Log}\frac{M}{M_\odot} =8.65\pm0.19  \\
   &{\rm Log}\frac{\dot M}{M_\odot/\rm s} =-8.14^{+0.16}_{-0.06} 
\end{split}
\end{equation} 

The confidence intervals reported here are defined as those values outside which the SED does not properly describe the dataset reliably. 
Specifically, we explored the $M$ and $\dot M$ parameter spaces to identify the limits for which the SED emission profile start to deviate significantly from both photometric and spectroscopic data. These limits define our confidence interval. 
Noticeably, the confidence interval on the black hole mass derived with this approach is smaller than the intrinsic uncertainty on the virial based $M$ estimation method ($\sim0.4$dex).
This first analysis suggests a sub-Eddington regime, with $\dot M/\dot M_{\rm Edd}=0.02\pm0.01$.

The analytic approximations we will use in Sections \ref{sec:kerr} and \ref{sec:slim} need the disk emission peak frequency and luminosity as input parameters. 
Therefore we derived these values for the best fitting SS73 model: 
\begin{equation}
\label{eq:peak}
\begin{split}
    \nu_{\rm max} &=2.01^{+0.60}_{-0.47}\times10^{15}\rm Hz &&&
    \nu L_{\rm max}(\nu) &=1.03^{+0.48}_{-0.13}\times10^{45}\rm erg/s
\end{split}\end{equation}

The confidence intervals on the peak position in frequency and luminosity are calculated as the peak values of the SED emission profiles obtained by assuming the confidence interval extremes of $M$ and $\dot M$ as input.

   \begin{figure}
   \centering
      \includegraphics[width=\hsize]{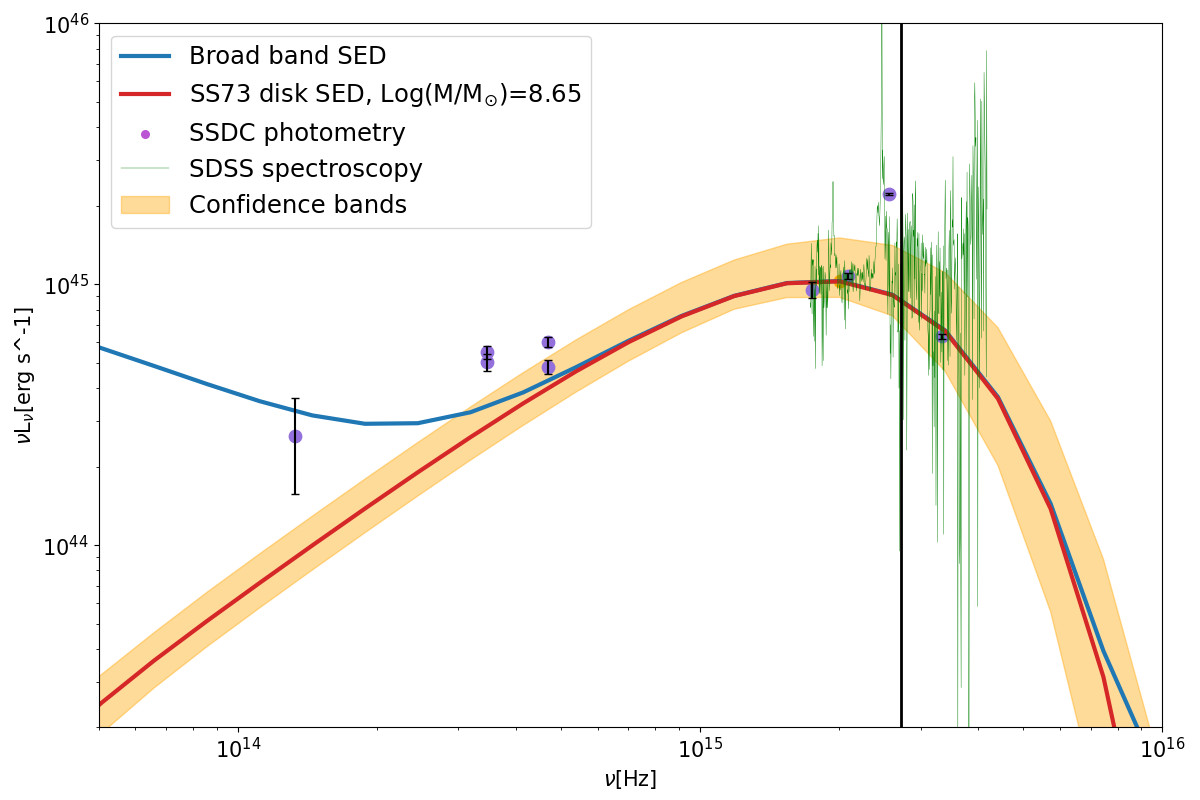}

      \caption{Zoomed-in visual of the most reliable model for \blaz, focussed on the accretion disk emission.
      As in Figure \ref{Fig:SED}, photometric and spectroscopic data are shown as purple points and green line, respectively, 
      while the red solid line shows the SS73 model obtained with a mass of $4\times10^8M_\odot$ and accretion rate of about 2\%.
      The black vertical line highlights the Ly$\alpha$ line frequency, above which we suspect that the \cite{meiksin06} algorithm over-correct the intervening absorption. Therefore, we do not consider that section of the spectrum in our modeling.
      }
         \label{Fig SS_SED}
   \end{figure}

\subsection{Considering a spinning black hole}
\label{sec:kerr}

SS73 has been widely used to model AGN emission, even in case of jetted sources and blazars in particular, but has a strong limitation: it does not include any effect linked to the black hole spin. 
Since jets in AGN are generally associated with a spinning SMBH, we decided to test a model that includes this parameter.
KERRBB \citep{KERRBB} describe the emission from a geometrically thin and optically thick accretion disk, including the General Relativity effects on dynamics and emission associated to a spinning black hole. KERRBB does not have a significantly different emission profile compared to SS73, specifically in their peak region. Therefore one can reproduce the other with the right set of physical parameters. We can thus use the peak position derived in Section \ref{sec:alpha} to apply the KERRBB analytic approximation built by \citet[][equations A.1, A.2 and A.3]{Campitiello_19}:
\begin{equation}
    \frac{M}{10^9M_\odot}=\left[\frac{g_1(a,\theta_\nu)\mathcal{A}}{\nu_p}\right]^2\sqrt{\frac{\nu_pL_p}{\mathcal{B}\cos\theta_\nu\, g_2(a,\theta_nu)}}
    \label{KERRBB_M}
\end{equation}
\begin{equation}
    \frac{\dot M}{M_\odot\rm yrs^{-1} }=\frac{\nu_pL_p}{\mathcal{B}\cos\theta_\nu\, g_2(a,\theta_nu)}
    \label{KERRBB_dotM}
\end{equation}
\begin{equation}
    \lambda=\mathcal{D}\frac{\eta(a)}{g_1^2(a,\theta_\nu)\sqrt{\cos\theta_\nu\,  g_2(a,\theta_\nu)}}\nu_p^2\sqrt{\nu_pL_p}
\end{equation}
where $\lambda=L_{\rm d}/L_{\rm Edd}$ is the luminosity-defined Eddington ratio, $\eta(a)$ is the radiative efficiency depending on the SMBH spin \citep[see e.g.][]{madau14}, and:
\begin{equation}
\label{eq_g}
\begin{split}
    g_i(a,\theta)&=\alpha_i+\beta_iy_1+\gamma_iy_2+\delta_iy_3+\epsilon_iy_4+\xi_iy_5+\iota_iy_6 \\
    y_n&=\log(n-a) \qquad\qquad i=1,2
\end{split}
\end{equation}
The parameters used in equation \ref{eq_g} are taken from Table 2 of \cite{Campitiello_18}, and are listed in Table \ref{KERRBB_table_parameters_g},  while
$a$ is the spin of the black hole. As for the anisotropy of the emission, we assumed that the disk was face on, hence $\theta\sim0$.
This analytical approximation provides a family of sets of physical parameters $M$ , $\dot M$ and $a$, that reproduce the SS73 emission profile fixed at specific peak coordinates $\nu_p$, $\nu_pL(\nu_p)$. We derived the family of results for a set of spin values:
\begin{equation}
\label{spin_values}
    a={-1 , 0    ,0.5, 0.6 , 0.7  , 0.8  , 0.85 , 0.9  , 0.95 ,0.9982}
\end{equation}
thus calculating the  corresponding mass $M$ and Eddington ratio $\lambda$. 
We followed a tighter sampling for values corresponding to a co-rotating accretion disk, since this set up is most likely associated to jet launching. The set of $M$, $\lambda$ estimates are shown in Figure \ref{Fig:riassuntiva} (orange line). These results outline a large black hole mass of ${\rm Log}\,M/M_\odot\sim8.2-8.5$, compatible with the virial mass measured in \cite{Shen2011}, and again a sub-Eddington accretion regime ($\sim0.03-0.06$). 
We also estimated the value of the Eddington ratio $\lambda_{vir}^s=0.034$ and spin $a=0.627$ corresponding to the virial mass value.

Our analysis with the KERRBB analytical approximation outlines the necessity of a model voted specifically to super-Eddington accretion rates, to further test the sub-Eddington behaviour suggested up to now. 

\subsection{Can it be super-Eddington?}
\label{sec:slim}

In the context of a black hole accreting at rates \dotM$\sim{\rm \dot M_{Edd}}$, a more accurate depiction of the accretion disk and its radiative emission is given by the SLIMBH model \citep{Abramowicz_1988,sadowski09}. In this model some of the initial assumptions of the $\alpha$-disk are relaxed to obtain a more suitable description of the accretion disk. In particular: 
\begin{itemize}
    \item  the accreting matter follows a sub-Keplerian rotation in most of the disk and a super-Keplerian rotation in the inner part of the disk, instead of following a Keplerian motion as in SS73;
    \item  the inner edge of the disk is closer to the black hole than the corresponding $r_{\rm in}=R_{\rm ISCO}$ for high accretion rates;
    \item  the energy dissipation in SLIMBH depends strongly on advection processes, other than radiative emission. This phenomenon changes the geometry from a thin to a \textit{slim} disk, and results in a lower radiative efficiency, with a consequently different emitted flux $F(\nu)$.
\end{itemize}
As for the case of KERRBB, the emission profile of SLIMBH is  a multi-temperature black body with a peak profile consistent with SS73. Therefore we adopted the same approach as in Section \ref{sec:kerr}, focussing on an analytical approximation of the model developed by \cite{Campitiello_19}:
\begin{equation}
    \frac{\nu_p}{[Hz]}=1.22\mathcal{A}\lambda^{1/4}\left[\frac{M}{10^9M_\odot}\right]^{-1/4}g_{1,s}(a,\theta_\nu,\lambda)
\end{equation}
\begin{equation}
    \frac{\nu_pL(\nu_p)}{[erg/s]}=2.21\mathcal{B}\lambda\left[\frac{M}{10^9M_\odot}\right]\cos(\theta_\nu)g_{2,s}(a,\theta_\nu,\lambda)
\end{equation}
These two equations yield:
\begin{equation}
    [\nu_pL(\nu_p)]^{1/4}\nu_p=\mathcal{E}[g_{2,s}(a,\theta_\nu, \lambda)\cos(\theta_\nu)]^{1/4}[g_{1,s}(a,\theta_\nu, \lambda)]\sqrt{\lambda}
    \label{eq_EddRatio}
\end{equation}
With the assumption that $\theta\sim0$, $g_1$ and $g_2$ can be expressed as:
\begin{equation}\begin{split}
    &g_{i,s}(a,\theta_\nu,\lambda)=\alpha_{i,s}+\beta_{i,s}y_1+\gamma_{i,s}(y_1)^2+\delta_{i,s}(y_1)^3+\epsilon_{i,s}(y_1)^4 
    \\ 
    &y_1=\log(1-a)
    \end{split}
    \label{slimbh_g}
\end{equation}
Differently from the KERRBB case, for this approximation the parameters $\alpha,\beta,\gamma,\delta,\epsilon$ depend on the Eddington ratio, and can be generalized as:
\begin{equation}
    \chi_{i,s}(\lambda)=\overline{a}+\overline{b}\lambda+\overline{c}(\lambda)^2+\overline{d}(\lambda)^3+\overline{e}(\lambda)^4+\overline{f}(\lambda)^5
\label{slimbh_chi}
\end{equation}
The parameters $\overline{a}, \overline{b}, \overline{c}, \overline{d}, \overline{e}, \overline{f}$ are listed in Table \ref{table_parameters_slimbh}, which is derived from Table A2 of \cite{Campitiello_19} in the case with viewing angle $\theta_\nu=0$.

\begin{figure}
  \centering
  \includegraphics[width=0.9\hsize]{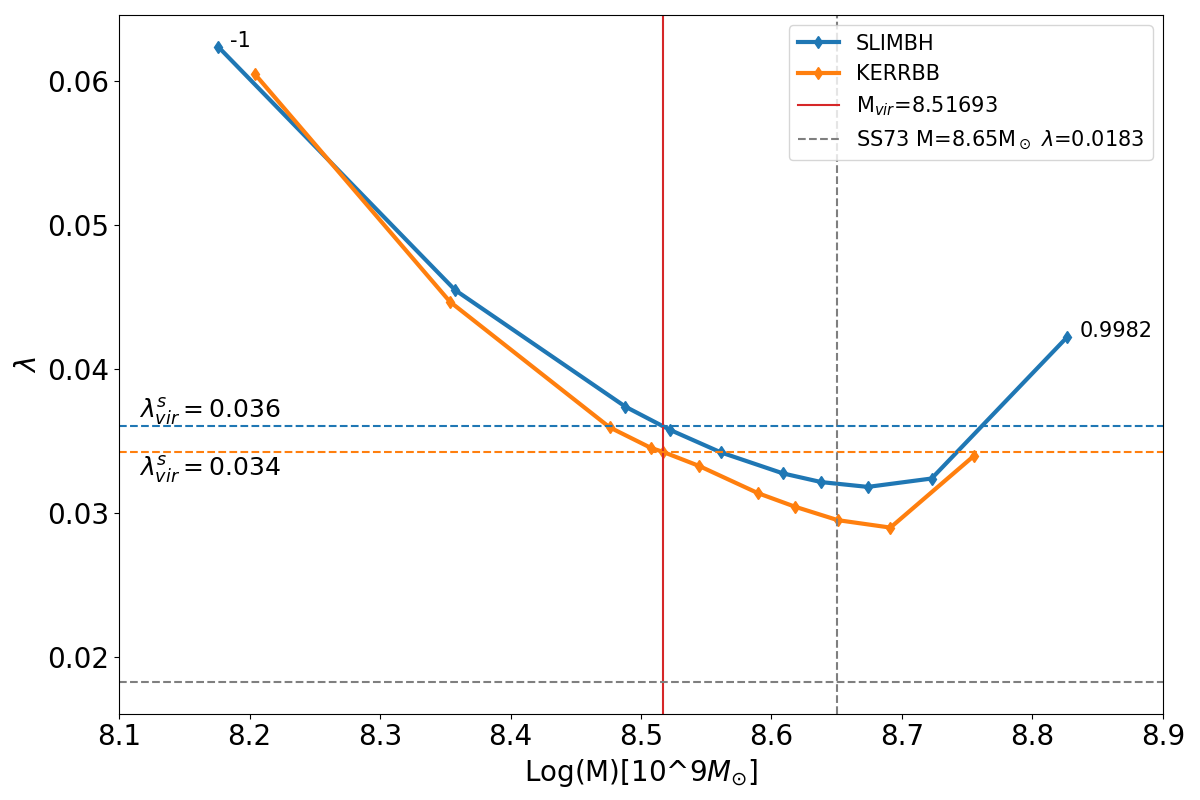}
  \caption{KERRBB (solid orange line) and SLIMBH (solid blue) sets of $M$ and $\lambda$ values consistent with \blaz\ data. 
  The two sets of results are compared with the virial mass (vertical red line) 
  in order to estimate the respective values of Eddington ratio $\lambda^s_{vir}$ (dashed blue and orange lines, for KERRBB and SLIMBH respectively). 
  The vertical and horizontal dashed grey lines show the estimates of $M$ and $\lambda$ derived with SS73, respectively.}
  \label{Fig:riassuntiva}
\end{figure}

Similarly to the analytical approximation of KERRBB, sets of three parameters $a$, $M$, $\lambda$ replicate the SS73 emission profile of given peak coordinates. We therefore followed an analogous procedure to what shown in Section \ref{sec:kerr} at the same fixed spin values (Equation \ref{spin_values}). Then we solved numerically Equation \ref{eq_EddRatio} to find the Eddington ratios and the masses. The resulting family of solutions is shown in Figure \ref{Fig:riassuntiva} (in blue) and show a large mass $8.18<{\rm Log}\,M/M_\odot<8.83$ and a sub-Eddington accretion rate $0.0320<\lambda<0.0624$. 
These values are consistent with what found following the KERRBB approximation. This is not surprising, given that SLIMBH at low accretion rates is consistent with the general relativistic assumptions included in KERRBB. 

The range of results obtained with our analysis draw indeed a strong conclusion: \blaz\ shows a sub-Eddington accretion regime even when a model specifically developed for close-to- or super-Eddington accretion is implemented.

\begin{table}
\caption{Physical parameters derived by modeling \blaz\ accretion disk with SS73, KERRBB and SLIMBH. 
Both KERRBB and SLIMBH provide range of parameters instead of specific best values, therefore here we include the maximally co- ($a=0.998$) and maximally counter-rotating ($a=-1.0$) solutions, along with the solutions corresponding to the independent virial-based mass measure by \cite[][i.e.\ $M_{\rm vir}=8.52 \pm 0.39$]{Shen2011}. 
}             
\label{table:1}      
\centering          
\begin{tabular}{l c c  c c}      
\hline\hline       
                      
\rule{0pt}{2.1ex}  
  & $a$& Log$\,M/M_\odot$ & $\eta$ & $\lambda$\\ 
\hline                    
\rule{0pt}{2.1ex}  
SS73 & 0& 8.6& 0.083 & 0.0183\\  
\hline
\rule{0pt}{2.1ex}  
KERRBB &-1 & 8.20    & 0.038 & 0.0605\\
  & 0.627 & 8.52     & 0.094 & 0.0342\\
  & 0.9982 & 8.76    & 0.329& 0.0340\\
\hline
\rule{0pt}{2.1ex}  
SLIMBH    & -1 & 8.18    & 0.038 & 0.0624\\
  & 0.5856 & 8.52     & 0.0897 & 0.0360\\
  & 0.9982 & 8.83    & 0.329& 0.0422 \\
\hline                  
\end{tabular}
\end{table}

\section{A mature jetted black hole in a young Universe}
\label{sec:salpeter}

According to all models tested, \blaz\ accretes at a significantly sub-Eddington accretion and hosts a massive black hole, yet these are not the most extreme values found for high-$z$ quasars. 
What does this imply for this source evolution history? 
If \blaz\ had accreted with the same estimated Eddington ratio since its seed formation, its seed black hole mass would not be compatible with the current theoretical expectations for black hole  formation in the early Universe (i.e.\ $\sim100M_\odot$). 

Assuming a Salpeter timescale \citep{Salpeter64} for its continuous accretion since $z\sim20$, we can estimate the black hole mass from which \blaz\ started its evolution. Depending on the assumed accretion disk model, the seed mass estimates slightly vary, but they are generally in a range $M_{\rm S}\sim2.5\times10^6-5\times10^8{\rm M_\odot}$, with a most likely value of $\sim1.3\times10^8{\rm M_\odot}$. The detailed values are reported in Table \ref{table:seed}. 
These results are clearly problematic according to seed formation theories \citep[see e.g.][]{Volonteri12}. 
In fact, even the  hypothesis of a massive gas cloud direct collapse into a black hole would not lead to such massive compact objects.

Clearly, the assumption of a continuous accretion with a fixed (small) Eddington ratio is not realistic. 
Simulations and theoretical models have lately suggested that high-$z$ quasars may have undergone super-Eddington accretion episodes during their evolution, to justify such large masses assembled in a short lifetime. 
The excess of jetted AGN observed at $z>4$ combined with the necessity of super-Eddington phases, suggest a link between the two factors: the jet might facilitate a faster accretion, exploiting part of the gravitational energy released during the accretion \citep{jolley08,ghisellini13}, or it might be linked to faster accretion episodes through different channels.

Since highly spinning black holes are expected to have a significant role in launching relativistic jets, one can expect that a merger or a continuous accretion episode could efficiently spin-up a black hole up to triggering the jet. 
After that, the accretion flow can slow down and stabilise on a more moderate accretion rate, while the black hole partially spins down.
In a simplified view, we will assume that \blaz\ has been accreting at the measured rate during its jet life, following an earlier super-Eddington phase. 
In the following, we will estimate the possible seed black hole values in this framework: a continuous super-Eddington accretion episode followed by a continuous accretion at the rate measured at $z=4.31$ for as long as the relativistic jet has been flowing. 

\blaz\ shows an extended jet in LOFAR images \citep{kappes22}. 
The authors estimated the age of this jetted structure by following various assumptions on its inclination and velocity. 
They derived an extension of 135 kpc, with a hot spot speed of $0.06c$, estimating an extended jet age of $\sim7.3$Myr. 
The jet results to be pretty young, not allowing for a significantly extended evolution with a standard accretion.
If we assume that \blaz\ have followed the observed accretion rate for the whole jet life, its central black hole must have been already 97-99\% of the mass measured at $z=4.31$.
We thus need to investigate how can this mass range be reached from a reasonable black hole seed ($5-200M_\odot$) before $z=4.329$ (i.e.\ 7.3Myr earlier than the measured redshift value for our source): we decide to explore an accretion regime limited to the Eddington luminosity, with two different assumptions on the radiative efficiency. 

Assuming an accretion with $\lambda\sim1$ implies a tight link between the matter accretion rate $\dot M/\dot M_{\rm Edd}$ and the radiative efficiency, since $\eta\,\dot M/\dot M_{\rm Edd}\sim1$. 
We assume (i) a ``standard" radiative efficiency of $\eta=0.1$, that implies an accretion rate $\dot M/\dot M_{\rm Edd}=10$, and (ii) a lower radiative efficiency of $\eta=0.05$, corresponding to $\dot M/\dot M_{\rm Edd}=20$, and approximating a much thinner optical depth, or a stronger photon trapping effect typical of highly accreting sources and/or highly spinning black holes. 
The choice of low radiative efficiencies is well motivated by literature, even when accretion occurs in a slim disk during the super-Eddington phase: the SMBH can be spun up by continuous coherent accretion, close to the maximally spinning value $a\sim1$, and yet have lower radiative efficiencies compared to the expected value $\sim0.3$, because of photon trapping effects \citep{thorne74}. 
If instead chaotic accretion dominates the Eddington or super-Eddington phases, the SMBH spin can be as low as $a\sim0$ \citep{dotti13}.
Note that these values are in agreement with the results presented by \cite{sadowski09} about the super-Eddington accretion through a slim disk.

Following these assumptions, the critical or super-critical accretion might have even started from a reasonable seed of $\sim50-200M_\odot$ already at redshift $z\sim8-5.5$. 
This is an extremely conservative result in terms of black hole seed formation: at $z\sim8$ massive PopIII stars have had enough time to evolve and collapse in a massive stellar black hole. 
We followed an approach that is still oversimplified, but can be easily applied to many massive jetted AGN in the early Universe, in order to put more reasonable constraints on their evolution, leading to better informed observational limit to specific theoretical frameworks and simulations. 

\begin{table}
\caption{Seed black hole mass values (last column) derived for different accreting scenarios, following SLIMBH results.
         The accretion histories are derived for a final black hole with mass and spin as reported in columns 1 and 2,
         continuously accreting with Eddington ratio and efficiency detailed in columns 3 and 4, starting from the redshift 
         shown in column 5. 
         The seeds located at $z=8$ and 5.5 are derived by assuming a continuous accretion up to the jet formation time ($z=4.329$). 
         }             
\label{table:seed}      
\centering                         
\begin{tabular}{c c c c c c}        
\hline\hline                 
\rule{0pt}{2.2ex}  
 Log$\,M/M_\odot$ & $a$ & $\lambda$ & $\eta$ & $z_{\rm seed}$ & Log$\,M_{\rm s}/M_\odot$ \\    
\hline
\rule{0pt}{2.5ex}  
  8.52 & 0.586 & 0.036 & 0.090 & 20.0 & 8.091 \\      
  8.52 & 0.586 & 0.036 & 0.090 & 4.329 & 8.515 \\      
  8.52 & 0.586 & 1.0 & 0.1 & 8.0 & 2.013 \\      
  8.52 & 0.586 & 1.0 & 0.5 & 5.5 & 2.027 \\      
\rule{0pt}{2.5ex}  
  8.83 & 0.998 & 0.042 & 0.329 & 20.0 & 8.730 \\      
  8.83 & 0.998 & 0.042 & 0.329 & 4.329 & 8.829 \\      
  8.83 & 0.998 & 1.0 & 0.1 & 8.0 & 2.328 \\      
  8.83 & 0.998 & 1.0 & 0.5 & 5.5 & 2.342 \\      
\rule{0pt}{2.5ex}  
  8.18 & -1.0 & 0.062 & 0.038 & 20.0 & 6.339 \\      
  8.18 & -1.0 & 0.062 & 0.038 & 4.329 & 8.169 \\      
  8.18 & -1.0 & 1.0 & 0.1 & 8.0 & 1.667 \\      
  8.18 & -1.0 & 1.0 & 0.5 & 5.5 & 1.681 \\      
\hline                                   
\end{tabular}
\end{table}

Another approach that allows for PopIII or massive stellar seed black holes at $z\sim20$ is the exploitation from the jet launching or accelerating mechanisms of part of the gravitational energy released during accretion \citep{jolley08}. \cite{ghisellini13} first suggested that the fraction of gravitational energy released $\eta_{\rm tot}$ can be splitted in the radiative efficiency $\eta$ that generates the observed accretion disk emission and an equivalent ``jet efficiency'' $\eta_{\rm j}$ that indicates the fraction of accretion energy involved in launching and/or accelerating the relativistic jet. They suggested this toy model in order to justify the presence of $z>4$, $M>10^9M_\odot$ blazars. 
This simple assumption would reconcile an effective energy release $\eta_{\rm tot}\sim0.3$ from the disk around a maximally spinning black hole, with a standard radiative efficiency $\eta\sim0.1$, leading to a much faster mass accretion on the SMBH for a fixed observed luminosity. 
This toy model could be applied in our case, too, and would certainly help in deriving a $\sim10^2M_\odot$ seed black hole at $z\sim20$ for \blaz, if the jet were active during the whole process. 
Nevertheless, the $\sim7$Myr old extended radio emission observed by LOFAR is not consistent with a jet age of about 1.3Gyr that would be needed for the hypothesis by \cite{ghisellini13}.
This is why in this work we explored a super-critical approach leading to jet triggering, instead of this elegant solution that showed its validity in the case of other $z>4$ blazars.

\begin{figure}
  \centering
  \includegraphics[width=1.1\hsize]{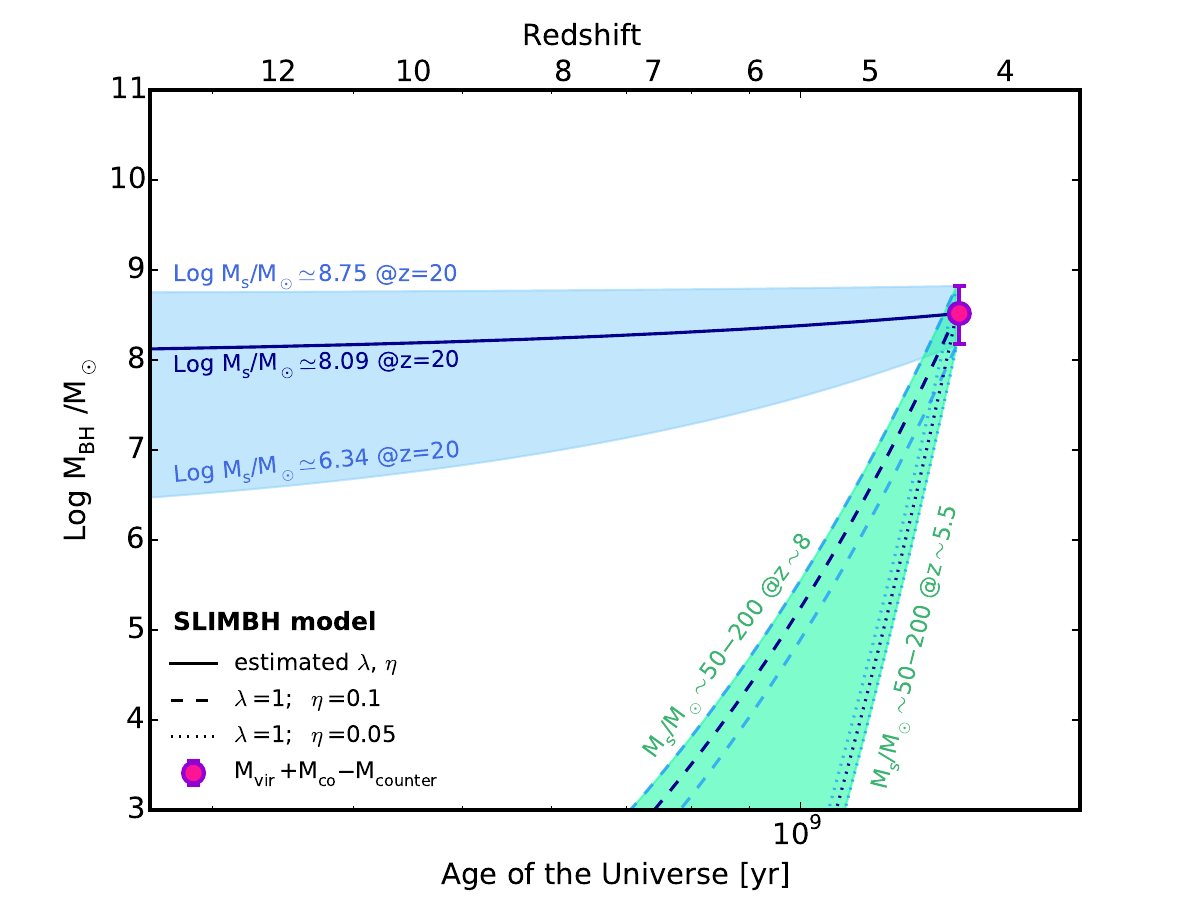}
  \caption{Black hole mass evolution as a function of time (or redshift, upper axis), 
           for the results obtained with the SLIMBH-based approach (KERRBB-based results are consistent). 
           The data point shows \blaz\ virial mass and the realtive mass range obtained following Section \ref{sec:slim}. 
           The solid blue line and light blue shaded area show how the mass would evolve 
           if it followed the same accretion regime as derived with SLIMBH since $z=20$. 
           The derived seed black hole masses $M_{\rm s}$ are labelled.
           The green shaded area instead follows an Eddington-limited evolution 
           in the case of a radiative efficiency that ranges between 5 and 10\% 
           (dotted and dashed lines, respectively). Following this accretion regime, 
           seed black holes in the range 50-200$M_\odot$ are inferred between $z=5.5$ 
           and $z=8$, respectively, as labelled.
           }
  \label{Fig:seed}
\end{figure}

\section{Conclusions}

In this paper we studied the broad-band emission of the $z=4.31$ {\it Fermi}-detected blazar \blaz, mainly focusing on its accretion regime and the nature of its seed black hole. 

\begin{itemize}
      \item[-] We first modelled the overall broad band SED with an analytic approximation of a standard blazar profile. 
      The broad-band SED is well constrained in its high-energy part, thanks to the $\gamma$-ray detection. 
      The strong Compton dominance is consistent with the blazars observed at high-$z$, typically more extreme than most {\it Fermi}-detected blazars. 
      \item[-] We analyzed the \blaz\ IR-optical-UV emission with a standard $\alpha$-disk model, varying its parameters to produce an emission profile that well described the photometric and spectroscopic data. 
      We derive a large black hole mass ${\rm Log}\,M/M_\odot=8.65\pm0.19$ and a sub-Eddington accretion rate $\lambda=0.02\pm0.01$.
      These results are consistent with the virial-based values obtained in literature. 
      \item[-] We then modelled the big blue bump with an analytic approximation of KERRBB developed by \cite{Campitiello_18}, that describes the emission from an accretion disk surrounding a spinning black hole, and takes into account general relativistic effects. 
      Even with this approach, \blaz\ results accreting significantly below the Eddington limit (${\rm Log}\,M/M_\odot\sim8.2-8.5$ and $\lambda\sim0.03-0.6$).
      \item [-] Finally, we tested the analytic approximation of SLIMBH developed by \cite{Campitiello_19}, that is similar to KERRBB but is extended at accretion rates larger than the Eddington limit. 
      Despite the possibility of exploring faster accretion regimes, \blaz\ once again results accreting at a sub-Eddington rate, consistent with the KERRBB-based results. 
\end{itemize}

The somewhat slow accretion rates obtained in this work are surprising in the context of very high redshift. 
Active SMBHs at $z>4$ are always found to be very massive, and the age of the Universe in this redshift range ($<1.5$Gyrs) is not enough for continuous, sub-critical, standard accretion to build them from even massive stellar seed black holes ($\sim10^2M_\odot$). 
We thus explored the possibility that \blaz\ had an early history of super-Eddington accretion, spinning up its black hole up to the point of launching a relativistic jet, and only after accreted at the observed rates. 
Having an estimate of the $\sim7.3$Myrs young jet age from LOFAR data \citep{kappes22}, we could reconcile the evolution of \blaz\ with black hole seeds of stellar mass. 

Once again, the  extreme accretion regimes that are expected to assemble the observed mass distribution of quasars and blazars at very high redshift are not caught in the act.
Nonetheless, every $z>4$ blazar detection or study suggests that relativistic jets might be involved in a fast accretion process, or at least be an effective tracer of a past super-critical event. 

\begin{acknowledgements}
    We would like to thank the referee, Markus Boettcher, for the useful suggestions that helped to clarify and improve the paper. 
    Part of this work is based on archival data and online services provided by the Space Science Data Center - ASI.
\end{acknowledgements}

\onecolumn
\begin{appendix}
\clearpage
\section{KERRBB and SLIMBH parameters}
\label{appendix}

In order to derive realistic estimates of mass, accretion rate and spin of \blaz, we applied two analytic approximations of numerical accretion models, i.e.\ KERRBB and SLIMBH as analysed by \citep{Campitiello_18,Campitiello_19}.
Tables \ref{KERRBB_table_parameters_g} and \ref{table_parameters_slimbh} report in detail the parameters used to build the SED emission profiles of face-on accretion disks according to the two models. 

\nopagebreak
\begin{table*}[!h]
    \centering
        \caption{Parameters used in Equation \ref{eq_g} to define $g_1$ and $g_2$ in the KERRBB analytic approximation, as derived by \cite{Campitiello_18} considering a viewing angle $\theta=0$ and reported in their Table 2. }
    \label{KERRBB_table_parameters_g}
    \begin{tabular}{c c c c c c c c}
    \hline
    \hline
    \rule{0pt}{2.5ex}
    Parameters   &  $\alpha_i$&$\beta_i$&$\gamma_i$&$\delta_i$&$\epsilon_i$&$\xi_i$&$\iota_i$ \\
    \hline
    \rule{0pt}{2.2ex}
        $g_1$&1001.3894&-0.061735&-381.64942&8282.077258&-40453.436&66860.08872&-34974.1536 \\
        $g_2$&2003.6451&-0.166612&-737.3402&16310.0596&-80127.1436&132803.23932&-69584.31533\\
        \hline
    \end{tabular}
\end{table*}

\begin{table*}[!h]
\centering
\caption{Parameters used in equation \ref{slimbh_chi} to define the SLIMBH analytic approximation, as derived by \citep{Campitiello_19} considering a viewing angle $\theta=0$. They correspond to $a,b,c,d,e,f$ in the original paper.}
\label{table_parameters_slimbh}
\begin{tabular}[]{c c c c c c c}
    \hline\hline
    \rule{0pt}{2.5ex}
    $\chi_{i,s}$& $\overline{a}$ &$ \overline{b}$ & $\overline{c}$&$ \overline{d}$&$ \overline{e}$& $\overline{f}$\\
    \hline
    \rule{0pt}{2.5ex}
        $\alpha_{1,s}$ & 1.37528& -0.03581&0.21458&-0.35971&0.16821&0\\
        $\beta_{1,s}$&-0.69241&0.21439&-0.57440&0.75878&-0.31345&0 \\
        $\gamma_{1,s}$&-0.28949&0.33854&-1.10213&1.27581&-0.48471&0\\
        $\delta_{1,s}$&-0.02094&0.20299&-0.65531&0.69134&-0.24475&0\\
        $\epsilon_{1,s}$&0.00538&0.04041&-0.12718&0.12852&-0.04411&0\\
    \hline
    \rule{0pt}{2.5ex}
        $\alpha_{1,s}$ &9.917034&-3.22512&17.89987&-41.84139&41.27971&-14.78658\\
        $\beta_{1,s}$&3.51987&-2.44127&13.36609&-29.61009&29.12455&-10.46037\\
        $\gamma_{1,s}$&0.72886&-4.09764&21.16717&-46.66724&46.53742&-16.81675\\
        $\delta_{1,s}$&0.39809&-2.70631&14.15261&-31.93365&32.12041&-11.63591\\
        $\epsilon_{1,s}$&0.10769&-0.52470&2.79254&-6.40072&6.46561&-2.34359\\
        \hline
    \end{tabular}
\end{table*}
\end{appendix}

\begin{thebibliography}{33}
\expandafter\ifx\csname natexlab\endcsname\relax\def\natexlab#1{#1}\fi

\bibitem[{{Abramowicz} {et~al.}(1988){Abramowicz}, {Czerny}, {Lasota}, \& {Szuszkiewicz}}]{Abramowicz_1988}
{Abramowicz}, M.~A., {Czerny}, B., {Lasota}, J.~P., \& {Szuszkiewicz}, E. 1988, \apj, 332, 646

\bibitem[{{Ackermann} {et~al.}(2017){Ackermann}, {Ajello}, {Baldini}, {Ballet}, {Barbiellini}, {Bastieri}, {Becerra Gonzalez}, {Bellazzini}, {Bissaldi}, {Blandford}, {Bloom}, {Bonino}, {Bottacini}, {Bregeon}, {Bruel}, {Buehler}, {Buson}, {Cameron}, {Caragiulo}, {Caraveo}, {Cavazzuti}, {Cecchi}, {Cheung}, {Chiang}, {Chiaro}, {Ciprini}, {Conrad}, {Costantin}, {Costanza}, {Cutini}, {D'Ammando}, {de Palma}, {Desiante}, {Digel}, {Di Lalla}, {Di Mauro}, {Di Venere}, {Dom{\'\i}nguez}, {Drell}, {Favuzzi}, {Fegan}, {Ferrara}, {Finke}, {Focke}, {Fukazawa}, {Funk}, {Fusco}, {Gargano}, {Gasparrini}, {Giglietto}, {Giordano}, {Giroletti}, {Green}, {Grenier}, {Guillemot}, {Guiriec}, {Hartmann}, {Hays}, {Horan}, {Jogler}, {J{\'o}hannesson}, {Johnson}, {Kuss}, {La Mura}, {Larsson}, {Latronico}, {Li}, {Longo}, {Loparco}, {Lovellette}, {Lubrano}, {Magill}, {Maldera}, {Manfreda}, {Marcotulli}, {Mazziotta}, {Michelson}, {Mirabal}, {Mitthumsiri}, {Mizuno}, {Monzani}, {Morselli}, {Moskalenko}, {Negro}, {Nuss}, {Ohsugi}, {Ojha},
  {Omodei}, {Orienti}, {Orlando}, {Ormes}, {Paliya}, {Paneque}, {Perkins}, {Persic}, {Pesce-Rollins}, {Piron}, {Porter}, {Principe}, {Rain{\`o}}, {Rando}, {Rani}, {Razzano}, {Razzaque}, {Reimer}, {Reimer}, {Romani}, {Sgr{\`o}}, {Simone}, {Siskind}, {Spada}, {Spandre}, {Spinelli}, {Stalin}, {Stawarz}, {Suson}, {Takahashi}, {Tanaka}, {Thayer}, {Thompson}, {Torres}, {Torresi}, {Tosti}, {Troja}, {Vianello}, \& {Wood}}]{ackermann17}
{Ackermann}, M., {Ajello}, M., {Baldini}, L., {et~al.} 2017, \apjl, 837, L5

\bibitem[{{Ajello} {et~al.}(2009){Ajello}, {Costamante}, {Sambruna}, {Gehrels}, {Chiang}, {Rau}, {Escala}, {Greiner}, {Tueller}, {Wall}, \& {Mushotzky}}]{ajello09}
{Ajello}, M., {Costamante}, L., {Sambruna}, R.~M., {et~al.} 2009, \apj, 699, 603

\bibitem[{{Atwood} {et~al.}(2009){Atwood}, {Abdo}, {Ackermann}, {Althouse}, {Anderson}, {Axelsson}, {Baldini}, {Ballet}, {Band}, {Barbiellini}, {Bartelt}, {Bastieri}, {Baughman}, {Bechtol}, {B{\'e}d{\'e}r{\`e}de}, {Bellardi}, {Bellazzini}, {Berenji}, {Bignami}, {Bisello}, {Bissaldi}, {Blandford}, {Bloom}, {Bogart}, {Bonamente}, {Bonnell}, {Borgland}, {Bouvier}, {Bregeon}, {Brez}, {Brigida}, {Bruel}, {Burnett}, {Busetto}, {Caliandro}, {Cameron}, {Caraveo}, {Carius}, {Carlson}, {Casandjian}, {Cavazzuti}, {Ceccanti}, {Cecchi}, {Charles}, {Chekhtman}, {Cheung}, {Chiang}, {Chipaux}, {Cillis}, {Ciprini}, {Claus}, {Cohen-Tanugi}, {Condamoor}, {Conrad}, {Corbet}, {Corucci}, {Costamante}, {Cutini}, {Davis}, {Decotigny}, {DeKlotz}, {Dermer}, {de Angelis}, {Digel}, {do Couto e Silva}, {Drell}, {Dubois}, {Dumora}, {Edmonds}, {Fabiani}, {Farnier}, {Favuzzi}, {Flath}, {Fleury}, {Focke}, {Funk}, {Fusco}, {Gargano}, {Gasparrini}, {Gehrels}, {Gentit}, {Germani}, {Giebels}, {Giglietto}, {Giommi}, {Giordano}, {Glanzman},
  {Godfrey}, {Grenier}, {Grondin}, {Grove}, {Guillemot}, {Guiriec}, {Haller}, {Harding}, {Hart}, {Hays}, {Healey}, {Hirayama}, {Hjalmarsdotter}, {Horn}, {Hughes}, {J{\'o}hannesson}, {Johansson}, {Johnson}, {Johnson}, {Johnson}, {Johnson}, {Kamae}, {Katagiri}, {Kataoka}, {Kavelaars}, {Kawai}, {Kelly}, {Kerr}, {Klamra}, {Kn{\"o}dlseder}, {Kocian}, {Komin}, {Kuehn}, {Kuss}, {Landriu}, {Latronico}, {Lee}, {Lee}, {Lemoine-Goumard}, {Lionetto}, {Longo}, {Loparco}, {Lott}, {Lovellette}, {Lubrano}, {Madejski}, {Makeev}, {Marangelli}, {Massai}, {Mazziotta}, {McEnery}, {Menon}, {Meurer}, {Michelson}, {Minuti}, {Mirizzi}, {Mitthumsiri}, {Mizuno}, {Moiseev}, {Monte}, {Monzani}, {Moretti}, {Morselli}, {Moskalenko}, {Murgia}, {Nakamori}, {Nishino}, {Nolan}, {Norris}, {Nuss}, {Ohno}, {Ohsugi}, {Omodei}, {Orlando}, {Ormes}, {Paccagnella}, {Paneque}, {Panetta}, {Parent}, {Pearce}, {Pepe}, {Perazzo}, {Pesce-Rollins}, {Picozza}, {Pieri}, {Pinchera}, {Piron}, {Porter}, {Poupard}, {Rain{\`o}}, {Rando}, {Rapposelli}, {Razzano},
  {Reimer}, {Reimer}, {Reposeur}, {Reyes}, {Ritz}, {Rochester}, {Rodriguez}, {Romani}, {Roth}, {Russell}, {Ryde}, {Sabatini}, {Sadrozinski}, {Sanchez}, {Sander}, {Sapozhnikov}, {Parkinson}, {Scargle}, {Schalk}, {Scolieri}, {Sgr{\`o}}, {Share}, {Shaw}, {Shimokawabe}, {Shrader}, {Sierpowska-Bartosik}, {Siskind}, {Smith}, {Smith}, {Spandre}, {Spinelli}, {Starck}, {Stephens}, {Strickman}, {Strong}, {Suson}, {Tajima}, {Takahashi}, {Takahashi}, {Tanaka}, {Tenze}, {Tether}, {Thayer}, {Thayer}, {Thompson}, {Tibaldo}, {Tibolla}, {Torres}, {Tosti}, {Tramacere}, {Turri}, {Usher}, {Vilchez}, {Vitale}, {Wang}, {Watters}, {Winer}, {Wood}, {Ylinen}, \& {Ziegler}}]{atwood09}
{Atwood}, W.~B., {Abdo}, A.~A., {Ackermann}, M., {et~al.} 2009, \apj, 697, 1071

\bibitem[{{Banados} {et~al.}(2024){Banados}, {Momjian}, {Connor}, {Belladitta}, {Decarli}, {Mazzucchelli}, {Venemans}, {Walter}, {Wang}, {Xie}, {Barth}, {Eilers}, {Fan}, {Khusanova}, {Schindler}, {Stern}, {Yang}, {Taufik Andika}, {Carilli}, {Farina}, {Fabian}, {Hennawi}, {Pensabene}, \& {Rojas-Ruiz}}]{banados24}
{Banados}, E., {Momjian}, E., {Connor}, T., {et~al.} 2024, arXiv e-prints, arXiv:2407.07236

\bibitem[{{Calderone} {et~al.}(2013){Calderone}, {Ghisellini}, {Colpi}, \& {Dotti}}]{calderone13}
{Calderone}, G., {Ghisellini}, G., {Colpi}, M., \& {Dotti}, M. 2013, \mnras, 431, 210

\bibitem[{{Campitiello} {et~al.}(2019){Campitiello}, {Celotti}, {Ghisellini}, \& {Sbarrato}}]{Campitiello_19}
{Campitiello}, S., {Celotti}, A., {Ghisellini}, G., \& {Sbarrato}, T. 2019, \aap, 625, A23

\bibitem[{{Campitiello} {et~al.}(2018){Campitiello}, {Ghisellini}, {Sbarrato}, \& {Calderone}}]{Campitiello_18}
{Campitiello}, S., {Ghisellini}, G., {Sbarrato}, T., \& {Calderone}, G. 2018, \aap, 612, A59

\bibitem[{{Dotti} {et~al.}(2013){Dotti}, {Colpi}, {Pallini}, {Perego}, \& {Volonteri}}]{dotti13}
{Dotti}, M., {Colpi}, M., {Pallini}, S., {Perego}, A., \& {Volonteri}, M. 2013, \apj, 762, 68

\bibitem[{{Fan} {et~al.}(2023){Fan}, {Ba{\~n}ados}, \& {Simcoe}}]{fan23}
{Fan}, X., {Ba{\~n}ados}, E., \& {Simcoe}, R.~A. 2023, \araa, 61, 373

\bibitem[{{Ghisellini} {et~al.}(2015){Ghisellini}, {Haardt}, {Ciardi}, {Sbarrato}, {Gallo}, {Tavecchio}, \& {Celotti}}]{ghisellini15}
{Ghisellini}, G., {Haardt}, F., {Ciardi}, B., {et~al.} 2015, \mnras, 452, 3457

\bibitem[{{Ghisellini} {et~al.}(2013){Ghisellini}, {Haardt}, {Della Ceca}, {Volonteri}, \& {Sbarrato}}]{ghisellini13}
{Ghisellini}, G., {Haardt}, F., {Della Ceca}, R., {Volonteri}, M., \& {Sbarrato}, T. 2013, \mnras, 432, 2818

\bibitem[{{Ghisellini} {et~al.}(2017){Ghisellini}, {Righi}, {Costamante}, \& {Tavecchio}}]{Ghisellini_Fermi_blazar_sequence}
{Ghisellini}, G., {Righi}, C., {Costamante}, L., \& {Tavecchio}, F. 2017, \mnras, 469, 255

\bibitem[{{Gokus} {et~al.}(2024){Gokus}, {B{\"o}ttcher}, {Errando}, {Kreter}, {He{\ss}d{\"o}rfer}, {Eppel}, {Kadler}, {Smith}, {Benke}, {Gurvits}, {Kraus}, {Lisakov}, {McBride}, {Ros}, {R{\"o}sch}, \& {Wilms}}]{gokus24}
{Gokus}, A., {B{\"o}ttcher}, M., {Errando}, M., {et~al.} 2024, \apj, 974, 38

\bibitem[{{Jolley} \& {Kuncic}(2008)}]{jolley08}
{Jolley}, E.~J.~D. \& {Kuncic}, Z. 2008, \mnras, 386, 989

\bibitem[{{Kappes} {et~al.}(2022){Kappes}, {Burd}, {Kadler}, {Ghisellini}, {Bonnassieux}, {Perucho}, {Br{\"u}ggen}, {Cheung}, {Ciardi}, {Gallo}, {Haardt}, {Morabito}, {Sbarrato}, {Drabent}, {Harwood}, {Jackson}, \& {Moldon}}]{kappes22}
{Kappes}, A., {Burd}, P.~R., {Kadler}, M., {et~al.} 2022, \aap, 663, A44

\bibitem[{{Kreter} {et~al.}(2020){Kreter}, {Gokus}, {Krauss}, {Kadler}, {Ojha}, {Buson}, {Wilms}, \& {B{\"o}ttcher}}]{kreter20}
{Kreter}, M., {Gokus}, A., {Krauss}, F., {et~al.} 2020, \apj, 903, 128

\bibitem[{{Li} {et~al.}(2005){Li}, {Zimmerman}, {Narayan}, \& {McClintock}}]{KERRBB}
{Li}, L.-X., {Zimmerman}, E.~R., {Narayan}, R., \& {McClintock}, J.~E. 2005, \apjs, 157, 335

\bibitem[{{Liao} {et~al.}(2018){Liao}, {Li}, \& {Fan}}]{liao18}
{Liao}, N.-H., {Li}, S., \& {Fan}, Y.-Z. 2018, \apjl, 865, L17

\bibitem[{{Liu} {et~al.}(2021){Liu}, {Wang}, {Momjian}, {Ba{\~n}ados}, {Zeimann}, {Willott}, {Matsuoka}, {Omont}, {Shao}, {Li}, \& {Li}}]{liu21}
{Liu}, Y., {Wang}, R., {Momjian}, E., {et~al.} 2021, \apj, 908, 124

\bibitem[{{Madau} {et~al.}(2014){Madau}, {Haardt}, \& {Dotti}}]{madau14}
{Madau}, P., {Haardt}, F., \& {Dotti}, M. 2014, \apjl, 784, L38

\bibitem[{{Meiksin}(2006)}]{meiksin06}
{Meiksin}, A. 2006, \mnras, 365, 807

\bibitem[{{Salpeter}(1964)}]{Salpeter64}
{Salpeter}, E.~E. 1964, \apj, 140, 796

\bibitem[{{Sbarrato}(2021)}]{sbarrato21}
{Sbarrato}, T. 2021, Galaxies, 9, 23

\bibitem[{{Sbarrato} {et~al.}(2012){Sbarrato}, {Ghisellini}, {Nardini}, {Tagliaferri}, {Foschini}, {Ghirlanda}, {Tavecchio}, {Greiner}, {Rau}, \& {Gehrels}}]{sbarrato12}
{Sbarrato}, T., {Ghisellini}, G., {Nardini}, M., {et~al.} 2012, \mnras, 426, L91

\bibitem[{{Sbarrato} {et~al.}(2015){Sbarrato}, {Ghisellini}, {Tagliaferri}, {Foschini}, {Nardini}, {Tavecchio}, \& {Gehrels}}]{sbarrato15}
{Sbarrato}, T., {Ghisellini}, G., {Tagliaferri}, G., {et~al.} 2015, \mnras, 446, 2483

\bibitem[{{Sbarrato} {et~al.}(2013){Sbarrato}, {Tagliaferri}, {Ghisellini}, {Perri}, {Puccetti}, {Balokovi{\'c}}, {Nardini}, {Stern}, {Boggs}, {Brandt}, {Christensen}, {Giommi}, {Greiner}, {Hailey}, {Harrison}, {Hovatta}, {Madejski}, {Rau}, {Schady}, {Sudilovsky}, {Urry}, \& {Zhang}}]{sbarrato13}
{Sbarrato}, T., {Tagliaferri}, G., {Ghisellini}, G., {et~al.} 2013, \apj, 777, 147

\bibitem[{{Shakura} \& {Sunyaev}(1973)}]{SS73}
{Shakura}, N.~I. \& {Sunyaev}, R.~A. 1973, \aap, 24, 337

\bibitem[{{Shen} {et~al.}(2011){Shen}, {Richards}, {Strauss}, {Hall}, {Schneider}, {Snedden}, {Bizyaev}, {Brewington}, {Malanushenko}, {Malanushenko}, {Oravetz}, {Pan}, \& {Simmons}}]{Shen2011}
{Shen}, Y., {Richards}, G.~T., {Strauss}, M.~A., {et~al.} 2011, \apjs, 194, 45

\bibitem[{{S{\k{a}}dowski}(2009)}]{sadowski09}
{S{\k{a}}dowski}, A. 2009, \apjs, 183, 171

\bibitem[{{Thorne}(1974)}]{thorne74}
{Thorne}, K.~S. 1974, \apj, 191, 507

\bibitem[{{Volonteri}(2012)}]{Volonteri12}
{Volonteri}, M. 2012, Science, 337, 544

\bibitem[{{Volonteri} {et~al.}(2011){Volonteri}, {Haardt}, {Ghisellini}, \& {Della Ceca}}]{Volonteri11}
{Volonteri}, M., {Haardt}, F., {Ghisellini}, G., \& {Della Ceca}, R. 2011, \mnras, 416, 216

\end{thebibliography}
\end{document}